\documentstyle[12pt]{article}
\newcommand{\beq}{\begin{equation}}
\newcommand{\eeq}{\end{equation}}
\newcommand{\bea}{\begin{eqnarray}}
\newcommand{\eea}{\end{eqnarray}}
\newcommand{\half}{\frac{1}{2}}
\newcommand{\ihalf}{\frac{i}{2}}

\newcommand{\G}{\Gamma}

\newcommand{\dslash}{\not\!\! D}

\newcommand{\psinapsi}{\bar{\psi}_{\rho}\hat{\nabla}_{\mu}\psi_{\sigma}}
\newcommand{\dmu}{\partial_{\mu}}

\renewcommand{\a}{\alpha}

\newcommand{\e}{\epsilon}
\renewcommand{\l}{\lambda}
\newcommand{\oz}{\bar{z}}
\newcommand{\ze}{\oz}

\newcommand{\opar}{\bar\epsilon}
\newcommand{\tres}{\;\;\;}
\begin{document}
\begin{titlepage}
\begin{flushright}
UB-ECM-PF 96/12 
\end{flushright}
\vspace{3mm}

\begin{center}

{\large {\bf Semi-local Cosmic Strings and the Cosmological Constant Problem}}
\vspace{1cm}

{\bf Jos\'e D. Edelstein}
\vspace{.5cm}

{\it Departament d'Estructura i Constituents de la Mat\`eria} \\
{\it Facultat de F\'{\i}sica, Universitat de Barcelona} \\
{\it Diagonal 647, 08028 Barcelona, Spain} \\
\vspace{.5mm}

{\it and}\\
\vspace{.5mm}

{\it Departamento de F\'{\i}sica, Universidad Nacional de La Plata} \\
{\it C.C. 67, (1900) La Plata, Argentina} \\ 
\vspace{1mm}

{\tt edels@venus.fisica.unlp.edu.ar}

\end{center}

\vspace{1.5cm}

\begin{abstract}
We study the cosmological constant problem in a 
three-di\-men\-sio\-nal $N=2$ supergravity theory with gauge group 
$SU(2)_{global} \times U(1)_{local}$. The model we consider is known to 
admit string-like configurations, the so-called semi-local cosmic strings.
We show that the stability of these solitonic solutions is provided by 
supersymmetry through the existence of a lower bound for the energy,
even though the manifold of the Higgs vacuum does not contain 
non-contractible loops. Charged Killing spinors do exist over 
configurations that saturate the Bogomol'nyi bound, as a consequence 
of an Aharonov-Bohm-like effect. 
Nevertheless, there are no physical fermionic zero modes on these 
backgrounds.
The exact vanishing of the cosmological constant does not imply, then, 
Bose-Fermi degeneracy.
This provides a non-trivial example of the recent claim made by Witten
on the vanishing of the cosmological constant in three dimensions without
unphysical degeneracies.
\end{abstract}

\end{titlepage}

The cosmological constant problem has longly survived the attempts made by
physicists to distangle it. One of the most interesting views on the
problem, in the context of three-dimensional supergravity, was recently
given by Witten \cite{Witten1}: supersymmetry can ensure the exact 
vanishing of the cosmological constant without compelling bosons and
fermions to be degenerate. Unbroken supercharges, which must be
constructed from spinors that are supercovariantly constant at infinity
are, in principle, ill-defined in $2+1$ as a consequence of the
asymptotically conical geometries produced by massive configurations
\cite{DJtH}. Then, even when supersymmetry applies to the vacuum ensuring
the vanishing of the cosmological constant, it is broken over the
excited states. Though there is no indication that an analogous scenario
can take place in four-dimensional space-times, some ways to extend this
result were thereafter explored, using the ideas of strong coupling
duality \cite{Witten2,Nishino} and S-duality \cite{KMS}.

The presumed non-existence of asymptotical Killing spinors in 
three-dimensional supergravities coupled to matter was recently
shown to be overcome in presence of Nielsen-Olesen vortices
\cite{BBS,ENS2}. The geometric phase associated with the conical
geometry results to be canceled by an Aharonov-Bohm phase produced by the
vortex flux lines. Nevertheless, the would-be fermionic Nambu-Goldstone
zero modes generated by the action of broken generators results to be
non-normalizable, thus not entering in the physical Hilbert space
\cite{BBS}. Then, in spite of being possible to end with Killing 
spinors over certain solitonic backgrounds, there is no Bose-Fermi 
degeneracy. It is still possible to have a vanishing cosmological 
constant without implying such an unphysical degeneracy in the 
spectrum.

This result was recently shown to apply, under general hypothesis, 
to any $2+1$ dimensional system admitting topological solitons \cite{ENS3}.
In all these models, the supercovariant derivative receives a contribution 
from the topological vector potential that allows the cancelation of
the conical holonomy by a phase produced when surrounding the solitonic
configuration.
Thus, it can be shown that the existence of non-trivial supercovariantly
constant spinors at infinity is guaranteed whenever massive field 
configurations saturate the corresponding Bogomol'nyi bound.
This assertion is still valid even if the topological vector potential
is an auxiliary field as in the $CP^n$ model discussed in \cite{ENS3}.
It seems to be a general result, however, that the fermionic zero modes 
receive an infrared divergent contribution from the gravitino transformation
law related to the conical nature of the $2+1$ dimensional
space-time. Then, vanishing of the cosmological constant can still be 
thought of as a consequence of an underlying supersymmetry, without
rendering bosons and fermions to be degenerate.

In the context of possible cosmic string scenarios, it is known that
stable solutions can take place even if the topology looks trivial in a
na\"{\i}ve sense: the manifold of minima for the potential energy does not
contain non-contractible loops. This is the case, for example, of the 
so-called semi-local cosmic string introduced in \cite{VA}, where stability 
is provided by the requirement that the gradient energy density falls off 
sufficiently fast at infinity. The stability of the static flat-space
string solution, as well as the critical relation between coupling 
constants where it takes place, can be understood as coming from
supersymmetry \cite{EN}: the semi-local cosmic string being thought
of as a purely bosonic configuration of an $N=2$ supersymmetric
system. In the gravitationally coupled system, semi-local cosmic string
and multi-string solutions were explicitely found and studied a few
years ago \cite{GORRS}. 

In the present letter we would like to address the would-be supersymmetric
nature of semi-local cosmic strings coupled to gravity, and its relation
with the cosmological constant problem. We show 
that the semi-local cosmic string can be consistently embedded into
an $N=2$ supergravity theory whenever the critical relation between 
coupling constants takes place. Its stability is shown to be a
consequence of the underlying $N=2$ supersymmetry algebra. We study
the existence of Killing spinors in a semi-local cosmic string
background, mainly focusing on the case in which it saturates a
Bogomol'nyi bound (though the system is, naively, non-topological).
We discuss the relation between these solutions and Witten's
claim about the vanishing of the cosmological constant without
Bose-Fermi degeneracy in $2+1$ dimensions.

Let us write down the $SU(2)_{global} \times U(1)_{local}$ lagrangian 
density of our $2+1$ dimensional system,
\beq
V^{-1}{\cal L} = \frac{M_{pl}}{2} R - \frac{1}{4} F_{\mu\nu}F^{\mu\nu}
+ \half (D_{\mu}\Phi)^{\dagger}(D^{\mu}\Phi) - \xi (\Phi^{\dagger}\Phi 
- v^2)^2 ~,
\label{modelo}
\eeq
which arises from the standard electroweak model minimally coupled
to gravity by setting the $SU(2)$ gauge coupling constant to zero.
$V$ is the determinant of the dreibein, $M_{pl}$ is the Planck
mass and $\xi$ is a real coupling constant.
The Higgs field $\Phi$ is a complex doublet
and the covariant derivative reflects the fact that we have only
gauged the $U(1)$ factor of the gauge group,
$D_{\mu} = \dmu - ieA_{\mu}$. The Lagrangian (\ref{modelo}) has
a global $SU(2)$ symmetry as well as a local $U(1)$ invariance,
under which the Higgs field changes as $\Phi \to e^{i\a}\Phi$.
When the Higgs field acquires a definite vacuum expectation value
the symmetry is broken to a global $U(1)$. The vacuum manifold is the
three-sphere $|\Phi| = v$, which has no non-contractible loops. 
However, as the gradient energy density must fall off sufficiently
fast asymptotically, fields at infinity owe to lie on a gauge
orbit, that is, a circle lying on the three-sphere.

We are, as announced, interested in cosmic string configurations 
of the system described by (\ref{modelo}) that could be understood as 
solutions of an $N=2$ supergravity theory. 
We must then accomodate our matter 
fields into an $N=2$ vector multiplet $(A_{\mu},\l,S)$ and an
$N=2$ hypermultiplet $(\Phi,\Psi)$, transforming in the vector representation
of $SU(2)$. We then accordingly enlarge the symmetry of our 
lagrangian density, coupling these multiplets to the Einstein
supermultiplet, finding, after some tensor calculus,
\bea
V^{-1}{\cal L}_{N=2} & = & \frac{M_{pl}}{2} R - \frac{1}{4} 
F_{\mu\nu}F^{\mu\nu} + \half (D_{\mu}\Phi)^{\dagger}(D^{\mu}\Phi) + 
\half \dmu S\partial^{\mu} S \nonumber \\
& + & \frac{e^2}{2}S^2\Phi^{\dagger}\Phi - 
\frac{e^2}{8}(\Phi^{\dagger}\Phi - v^2)^2 - 
\frac{V^{-1}}{2}\epsilon^{\rho\mu\sigma}\psinapsi \nonumber \\
& - & \half\bar{\lambda}\gamma^{\mu}\tilde{\nabla}_{\mu}\lambda
- \half\bar{\Psi}\gamma^{\mu}\tilde{\nabla}_{\mu}\Psi + 
V^{-1}{L}_{Fer}^{int} ~,
\label{lagn2}
\eea
where the last term ${L}_{Fer}^{int}$ includes several interaction terms
involving fermions, whose explicit form will not be of interest for us. 
Note that $N=2$ supersymmetry has forced a definite relation between
the Higgs coupling constant and the electric charge
\beq
\xi = \frac{e^2}{8} ~,
\label{cond}
\eeq
as it happens in the abelian Higgs model both in flat and curved 
space-time \cite{ENS2,ENS1}.
Concerning the fermion derivatives in (\ref{lagn2}), they are given by
the following expressions:
\beq
\hat{\nabla}_{\mu}\psi_{\sigma} = \left({\cal D}_{\mu} +
\frac{i}{4M_{pl}}J_{\mu} + ie\frac{v^2}{4M_{pl}}
A_{\mu}\right)\psi_{\sigma} ~,
\label{sup1}
\eeq
\beq
\tilde{\nabla}_{\mu}\lambda = \left({\cal D}_{\mu} - 
ie\frac{v^2}{4M_{pl}} A_{\mu}\right)\lambda ~,
\label{sup2}
\eeq
\beq
\tilde{\nabla}_{\mu}\Psi = \left({\cal D}_{\mu} - ie \left( 1 + 
\frac{v^2}{4M_{pl}}\right) A_{\mu}\right)\Psi ~,
\label{sup3}
\eeq
where ${\cal D}_{\mu}$ is the general relativity covariant derivative 
acting on a spinor, and $J_{\mu}$ is the Higgs field current,
\beq
J_{\mu} = \ihalf \left( (D_{\mu}\Phi)^{\dagger}\Phi - 
\Phi^{\dagger}(D_{\mu}\Phi) \right) ~.
\label{current}
\eeq
Notice that all fermions have received a charge contribution of magnitude
$\frac{ev^2}{4M_{pl}}$, as a consequence of the coupling of
our system to $N=2$ supergravity. This striking fact, related to the
presence of a Fayet-Iliopoulos term in the supergravity lagrangian,
induces a charged supersymmetric parameter. The relevance of this 
issue will become clear when studying the evasion of the no-go scenario
posed by the asymptotically conical nature of the $2+1$ dimensional
space-time.
We will be mainly interested in purely bosonic field 
configurations of this system. We then introduce the following useful
notation: given a functional $\Xi$ depending both on bosonic and 
fermionic fields, we will use $\Xi\vert$ to refer to that functional 
evaluated in the purely bosonic background,
\beq
\Xi\vert \equiv \Xi\vert_{\l,\Psi,\psi_{\mu}=0} ~.
\label{spb}
\eeq
Under this condition, the only non-trivial supersymmetric transformations
laws that leave invariant the lagrangian (\ref{lagn2}),
are those corresponding to fermionic fields:
\beq
\delta\lambda\vert = \half F_{\mu\nu}\sigma^{\mu\nu}\e +
\frac{ie}{4}(\Phi^{\dagger}\Phi - v^2)\e + \gamma^{\mu}\dmu S\e ~,
\label{traf1}
\eeq
\beq
\delta\Psi\vert = \half \left(\dslash\Phi + ieS\Phi\right) \e \tres ,
\tres \delta\psi_{\mu}\vert = 2M^{1/2}_{pl}\hat{\nabla}_{\mu}\e ~.
\label{traf2}
\eeq

We are interested in field configurations describing semi-local
cosmic strings that emerge from the system described by (\ref{modelo}). 
Thus, we shall make at this point $S = 0$.
Moreover, since Bogomol'nyi equations correspond to static
configurations with $A_0 = 0$, we also impose these conditions
(note that in this case $T_{0i}^{mat} = 0$).

The metric of a static spacetime can always be adapted to the
time-like killing vector field $\frac{\partial}{\partial t}$, such
that it is given by: 
\beq
ds^2 = H^2dt^2 - \Omega^2dzd\ze ~,
\label{metric}
\eeq
where we have written the metric on the surface $\Pi$, orthogonal 
everywhere to $\frac{\partial}{\partial t}$, in terms of a K\"ahler 
form and complex local coordinates, $\Omega$ being the conformal 
factor. Functions $H$ and $\Omega$ depend only on complex coordinates
$z$ and $\ze$. 
Far from the finite-energy matter sources, it is well-known that the
metric  must approach a cone with deficit angle $\delta$, whose 
explicit value will be derived below. This behaviour can be expressed
in terms of the following asymptotic conditions for functions
$H$ and $\Omega$,
\beq
H \to 1 \tres , \tres \Omega \to |z|^{-\delta/\pi} ~.
\label{metricbis}
\eeq
With this metric, the only non-vanishing components of the
Einstein tensor $G_{\mu\nu} = R_{\mu\nu} - \half g_{\mu\nu}R$, are
\beq
G_{00} = H^2K - H\vec\nabla^2H ~,
\label{einstens1}
\eeq
\beq
G_{ij} = -2H^{-1}\nabla_i\nabla_jH + k_{ij}H^{-2}(\vec\nabla H)^2 +
2k_{ij}H^{-1}\vec\nabla^2H ~,
\label{einstens2}
\eeq
where $K$ is the Gauss curvature of the two-dimensional metric $k_{ij}$
that spans $\Pi$, while $\nabla_i$ is the covariant derivative with 
respect to the planar metric. The integral of the Gauss curvature over
the surface $\Pi$ can be evaluated using the Gauss-Bonnet theorem:
\beq
\int_{\Pi} dzd\ze \Omega^2 K = \delta - 4\pi g ~,
\label{gabon}
\eeq
where $\delta$ is the deficit angle and $g$ is the number of handles
of the two-dimensional manifold $\Pi$. 
We will be concerned with the simplest case where the topology of $\Pi$
is that of a two-disk, $g = 0$. 
The $00$-component of the Einstein equations $G_{\mu\nu} = 
\kappa^2T_{\mu\nu}^{~mat}$, can be the used to obtain an explicit
equation for the deficit angle,
\beq
\delta = \int_{\Pi} dzd\ze \Omega^2 \left[\left(\vec\nabla\ln{H}\right)^2
+ \frac{1}{M_{pl}} T^{0 mat}_{~0}\right] \geq \frac{M}{M_{pl}} ~,
\label{delta}
\eeq
which results to be bounded by the total mass of the field 
configuration,
\beq
M = \half \int dzd\ze \left(\frac{\Omega}{H}\right)^2 
\left[F_{z\ze}F^{z\ze} + (D_{i}\Phi)^{\dagger}(D^{i}\Phi) + 
\frac{e^2}{4}(\Phi^{\dagger}\Phi - v^2)^2\right].
\label{muno}
\eeq
The bound is saturated provided
\beq
\nabla_i\ln{H} = 0 ~,
\label{cerocom}
\eeq
holds locally, this implying that $H$ is a constant which, after
conditions (\ref{metricbis}), is taken to be one, $H = 1$.

As we have previously mentioned, the existence of a spinor that is
asymptotically supercovariantly constant is related to its properties
under parallel transport at infinity. Consider, for example, in pure
supergravity, a spinor $\eta$ with definite `chirality', 
$\gamma^0\eta_{\pm} = \pm\eta_{\pm}$. Then, the parallel transport
around a closed curve $\Gamma$ of large radius $R$, surrounding all
the static matter sources, is given by the following path-ordered 
integration: 
\bea
\eta_{\pm}(R,2\pi) & = & {\cal P}\exp\left(-\ihalf\oint_{\Gamma}
\omega_{\mu}^a\gamma_adx^{\mu}\right)\eta_{\pm}(R,0) \nonumber \\
& = & \exp\left[\pm i\pi\delta\right]\eta_{\pm}(R,0) ~.
\label{paral}              
\eea
In view of (\ref{delta}), a non-trivial holonomy of geometric
nature arises for masses below the Planck scale, this resulting in an 
ill-defined spinor.

In our model, however, the supercovariant derivative receives corrections
from the fact that the gravitino is charged under the gauged $U(1)$
symmetry. Then, the previous argument slightly modifies to give:
\bea
\eta_{\pm}(R,2\pi) & = & {\cal P}\exp\left(- \ihalf\oint_{\Gamma}
\omega_{\mu}^a\gamma_adx^{\mu} + \frac{i}{4M_{pl}}\oint_{\Gamma}(J_{\mu} 
+  ev^2A_{\mu})dx^{\mu}\right)\eta_{\pm}(R,0) \nonumber \\
& = & {\cal P}\exp\left(- \ihalf\oint_{\Gamma}
\omega_{\mu}^a\gamma_adx^{\mu} + i\frac{ev^2}{4M_{pl}}
\oint_{\Gamma}A_{\mu}dx^{\mu}\right)\eta_{\pm}(R,0) ~,
\label{pax}
\eea
where we have imposed the constraint of finite energy on the Higgs field
current. We inmediately see that the charged spinor $\eta$, acquires
a Aharonov-Bohm phase provided some magnetic flux $\Phi_n$ exists across 
the surface delimited by $\G$. Indeed, eq.(\ref{pax}) can be rewritten as
\beq
\eta_{\pm}(R,2\pi) = \exp\left[ \pm i\pi \left( \delta \pm 
\frac{M_v^2}{4eM_{pl}}\Phi_n \right) \right]\eta_{\pm}(R,0) ~,
\label{pax2}
\eeq
where $M_v^2 = e^2v^2$ is the `photon' mass and $\Phi_n$, as discussed
above, is quantized:
\beq
\Phi_n = - \frac{2\pi n}{e} ~.
\label{cuant}
\eeq
There are vortex configurations that solve the Bogomol'nyi equations
of the system (first-order differential equations whose solutions also
satisfy the more involved Euler-Lagrange ones), for which phases 
exactly cancel \cite{GORRS}. Then, asymptotical Killing spinors do
exist over these $2+1$-dimensional background solutions and their 
corresponding unbroken supercharges should be built. 
Let us analyse more carefully this issue in order to see whether it 
leads or not to an evasion of Witten's
claim on the cosmological constant problem. We first construct the 
supercharge algebra to have a deeper understanding on its connection
with the Bogomol'nyi bound of the system.

In analogy to what happens in four-dimensional theories \cite{T,BCT},
the supercharge in three-dimensional supergravity is given by a surface
integral
\beq
{\cal Q}[\epsilon] = 2M^{1/2}_{pl}
\oint_{\partial\Sigma}\opar\psi_{\mu}dx^{\mu} ~,
\label{qsurf}
\eeq
which is nothing but the circulation of the gravitino
arround the oriented boundary $\partial\Sigma$ of a space-like surface
$\Sigma$ \cite{ENS2}. This quantity can not be used naively as a 
generator taking its Poisson brackets with a given field. One should
first fix the whole gauge freedom such that the asymptotic value of the
supersymmetry parameter determines its value everywhere \cite{T,DT}.
Alternatively, we can compute the algebra just by acting on the integrand
of (\ref{qsurf}) with a supersymmetry transformation:
\beq
\{\bar{{\cal Q}}[\epsilon],{\cal Q}[\epsilon]\}\vert \equiv
2M^{1/2}_{pl} \oint_{\partial\Sigma}\opar\delta_{\e}\psi_{\mu} dx^{\mu}
= 4M_{pl} \oint_{\partial\Sigma}\opar\hat{\nabla}_{\mu}\e dx^{\mu} ~.
\label{qeqe}
\eeq
This expression relates, as usual, the supercharge algebra evaluated in 
the purely bosonic sector with the circulation of a generalized Nester 
form $\opar\hat{\nabla}\e$. We now impose a chirality 
condition over the spinor $\gamma^0\e_{\pm} = \pm\e_{\pm}$, and choose
appropriate asymptotic conditions on the fields (i.e. consistent with
finite energy recquirements). In order to compute the integral, we 
further consider the contour of integration at large but finite radius 
$R$ (to avoid infrared problems). Then, for static configurations, it is
straightforward to obtain the result:
\beq
\{\bar{{\cal Q}}[\e_{\pm}],{\cal Q}[\e_{\pm}]\}\vert = 4\pi M_{pl} 
\left( \delta \pm \frac{M_v^2}{4eM_{pl}}\Phi_n \right)
\e^{\dagger}_{\pm\infty}\e_{\pm\infty}\Theta^{2}(R) ~,
\label{lns}
\eeq
where $\Theta(R)$ gives the asymptoptic behaviour of the supersymmetry
parameter 
\beq
\e_{\pm} \rightarrow \Theta(R)\e_{\pm\infty} ~.
\label{compasimp}
\eeq
One can also compute the supercharge algebra in a different (longer but
more illuminating) way. We can use Stokes' theorem to rewrite the 
r.h.s. of (\ref{qeqe}) as
\beq
\{\bar{{\cal Q}}[\e_{\pm}],{\cal Q}[\e_{\pm}]\}\vert = 2M_{pl} 
\int_{\Sigma}\epsilon^{\mu\nu\beta}\nabla_{\beta}
(\opar_{\pm}\hat{\nabla}_{\mu}\e_{\pm})d\Sigma_{\nu} ~.
\label{cadorna}
\eeq
Then, as the commutator of supercovariant derivatives is given by:
\beq
[\hat{\nabla}_{\mu},\hat{\nabla}_{\nu}] = 
\half R_{\mu\nu}^{~~a}\gamma_{a}
+ \frac{i}{4M_{pl}} (\partial_{\mu}J_{\nu} - \partial_{\nu}J_{\mu}
+ ev^2F_{\mu\nu}) ~,
\label{nmunu}
\eeq
after using Einstein equations and the explicit form of the supersymmetry 
transformation laws of the fermionic fields (\ref{traf1})-(\ref{traf2}), 
we obtain:
\bea
\{\bar{{\cal Q}}[\e_{\pm}],{\cal Q}[\e_{\pm}]\}\vert & = & 2M_{pl}
\int_{\Pi} dzd\ze\Omega^2 \left[ 
\left(\not\!\hat{\nabla}\e_{\pm}\right)^{\dag}
\left(\not\!\hat{\nabla}\e_{\pm}\right) - 
\left(\hat{\nabla}_i\e_{\pm}\right)^{\dag}
\left(\hat{\nabla}^i\e_{\pm}\right) \right. \nonumber \\
& + & \left. \frac{1}{2M_{pl}} \left(
\delta_{\e_{\pm}}\Psi\vert^{\dag}\delta_{\e_{\pm}}\Psi\vert +
\delta_{\e_{\pm}}\l\vert^{\dag}\delta_{\e_{\pm}}\l\vert \right) 
\right] ~,
\label{compcero}
\eea
where we have specialized our spacelike integration surface $\Sigma$ so 
that $d\Sigma_{\mu} = (d\Sigma_{0},\vec{0})$.
At this point, we note that after imposing a generalized Witten 
condition \cite{W} on the spinorial parameter $\epsilon$
\beq 
\gamma^i\hat{\nabla}_i\e_{\pm} = 0,
\label{genw}
\eeq
the asymptotic behaviour of $\epsilon$ can be determined and the r.h.s 
of eq.(\ref{compcero}) results to be 
a sum of bilinear terms hence semi-positive definite\footnote{It is 
natural to impose such a condition in order to consistently eliminate
gauge degrees of freedom. In fact, unphysical excitations are eliminated
in the transverse gauge $\gamma^i\psi_i = 0$, and this gauge fixing
is respected by supersymmetry provided $\e$ satisfies the generalized
Witten condition (\ref{genw}).}:
\beq
\{\bar{{\cal Q}}[\epsilon],{\cal Q}[\epsilon]\}\vert \geq 0 ~.
\label{supbound}
\eeq
In view of (\ref{lns}), this inequality corresponds to a bound on the
deficit angle
\beq
\delta \geq \mp \frac{M_v^2}{4eM_{pl}} \Phi_n ~,
\label{defibound}
\eeq
which results to be saturated if and only if
\beq
\delta_{\e_{\pm}}\Psi = \delta_{\e_{\pm}}\l = 0 ~,
\label{bogomat}
\eeq
\beq
\delta_{\e_{\pm}}\psi_{\mu} = 0 ~.
\label{bogograv}
\eeq
We recognize in the first couple of equations the Bogomol'nyi equations
for matter fields
\beq
{\cal F} = \frac{e}{2}(\Phi^{\dagger}\Phi - v^2) = 0 \tres , \tres
D_z\Phi = 0 ~,
\label{bogomas}
\eeq
for the upper sign, and
\beq
{\cal F} = - \frac{e}{2}(\Phi^{\dagger}\Phi - v^2) = 0 \tres , \tres
D_{\ze}\Phi = 0 ~.
\label{bogomen}
\eeq
for the lower one. Here ${\cal F} = \e^{z\ze}F_{z\ze}$, where
$\e^{z\ze}$ is the covariant antisymmetric tensor.
These equations were originally found in Ref.\cite{GORRS}
for the bosonic system in a non-supersymmetric context.
Let us remark that we have obtained them just by asking
our configuration to have an unbroken supersymmetry or,
in other words, to saturate the lower bound that results
from the supercharge algebra. Concerning equation (\ref{bogograv}), its 
solvability can be studied from the integrability conditions
\beq
[\hat{\nabla}_{\mu},\hat{\nabla}_{\nu}]\e_{\pm} = 0 ~,
\label{integral}
\eeq
which happen to be equivalent to the Einstein equations of the purely
bosonic system, once Bogomol'nyi equations (\ref{bogomas}) or
(\ref{bogomen}), according to the chirality of $\e$, have been imposed.
This leaves us with an exact Killing spinor preserved by this 
$2+1$-dimensional solitonic background. Even though a na\"{\i}ve
analysis of the vacuum manifold of this system would lead us to
conclude that it has a trivial topology, we have shown that the no-go 
scenario for the existence of Killing spinors in an asymptotically conical 
space-time is indeed overcome. This further extends the class of 
$2+1$-dimensional systems studied in \cite{ENS3}, that admits Killing 
spinors over massive configurations.

Let us end this letter by considering the connection between the
unbroken supersymmetry found above, and the cosmological constant
problem in the context of supergravity theories. In order to attempt the
construction of the entire  massive supermultiplet associated to a given 
semi-local cosmic string (that saturates the Bogomol'nyi bound), we must 
apply the broken supersymmetry generator over our purely bosonic 
configuration. Let us assume that the Killing spinor has chirality 
$\e_+$ and equations (\ref{bogomas}) hold, then:
\beq
\delta_{\e_-}\l\vert = \frac{ie}{2}(\Phi^{\dagger}\Phi - v^2)\e_- \neq 0
\tres , \tres \delta_{\e_-}\Psi\vert = \half\gamma^{\ze}D_{\ze}\Phi\e_- 
\neq 0 ~,
\label{zerouno}
\eeq
\beq
\delta_{\e_-}\psi_{\mu}\vert = 2M^{1/2}_{pl}\hat{\nabla}_{\mu}\e_- \neq 0 ~,
\label{zerodos}
\eeq
are nothing but the Nambu-Goldstone fermionic zero mode corresponding to
the unbroken supersymmetry. We note that the asymptotic behaviour of $\e_-$ 
given in (\ref{compasimp}) leads, for masses below the Planck scale,
to an infrared divergent contribution coming from eq.(\ref{zerodos}),
that renders the zero mode non-normalizable. As a consequence, the zero 
mode must not be used to construct the physical Hilbert space of the 
theory. That is, though there seem to exist unbroken supersymmetries
over certain massive configurations, they cannot be realized on the
physical spectrum. Hence, in this model, the vanishing of the cosmological
constant implied by the supersymmetries of the vacuum, does not compell
bosons and fermions to be degenerate. It would be very interesting to
generalize this mechanism to $3+1$-dimensional systems. We hope to report 
on this issue in a forthcoming work.

This work has been partially supported by CONICET.
I would like to thank the International Center for Theoretical Physics 
for its kind hospitality during the last stage of this work.


\begin{thebibliography}{99}
\bibitem{Witten1} E. Witten, Int. J. Mod. Phys. {\bf A 10} (1995) 1247.
\bibitem{DJtH} S. Deser, R. Jackiw y G. 't Hooft, Ann. Phys. (N.Y.)
{\bf 152} (1984) 220.
\bibitem{Witten2} E. Witten, Mod. Phys. Lett. {\bf A10} (1995) 2153.
\bibitem{Nishino} H. Nishino, Phys. Lett. {\bf B370} (1996) 65.
\bibitem{KMS} S. Kar, J. Maharana and H. Singh, Phys. Lett. {\bf B374}
(1996) 43.
\bibitem{BBS} K. Becker, M. Becker and A. Strominger, Phys. Rev. {\bf D51}
(1995) R6603.
\bibitem{ENS2} J.D. Edelstein, C. N\'u\~nez and F.A. Schaposnik,
Nucl. Phys. {\bf B458} (1996) 165.
\bibitem{ENS3} J.D. Edelstein, C. N\'u\~nez and F.A. Schaposnik,
Phys. Lett. {\bf B375} (1996) 163.
\bibitem{VA} T. Vachaspati and A. Ach\'ucarro, Phys. Rev. {\bf D44} 
(1991) 3067.
\bibitem{EN} J.D. Edelstein and C. N\'u\~nez, {\em Supersymmetric Electroweak 
Cosmic Strings}, preprint UB-ECM-PF 96/9 and La Plata-Th 96/07, 
hep-th/9605066.
\bibitem{GORRS} G.W. Gibbons, M.E. Ortiz, F. Ruiz Ruiz and T.M. Samols,
Nucl. Phys. {\bf B385} (1992) 127.
\bibitem{ENS1} J.D. Edelstein, C. N\'u\~nez and F.A. Schaposnik,
Phys. Lett. {\bf B329} (1994) 39.
\bibitem{T} C. Teitelboim, Phys.Lett. {\bf B 69} (1977) 240.
\bibitem{BCT} R. Benguria, P. Cordero and C. Teitelboim, Nucl.Phys.
{\bf B122} (1977) 61.
\bibitem{DT} S. Deser and C. Teitelboim, Phys. Rev. Lett. {\bf 39} (1977) 249.
\bibitem{W} E. Witten, Comm. Math. Phys. {\bf 80} (1981) 381.
\end{thebibliography}
\end{document}